\documentclass[aps,prl,twocolumn,psfig,showpacs]{revtex4}
\usepackage{amsfonts}
\usepackage{amsmath}
\usepackage{graphicx}
\usepackage{bm}
\usepackage{amssymb}
\usepackage{times}
\usepackage{dcolumn}
\usepackage{cases}
\usepackage{txfonts}

\begin{document}

\title{Linearized Tensor Renormalization Group Algorithm for Thermodynamics of Quantum Lattice Models}
\author{Wei Li$^{1}$, Shi-Ju Ran$^{1}$, Shou-Shu Gong$^{1}$, Yang Zhao$^{1}$%
, Bin Xi$^{1}$, Fei Ye$^{2}$, and Gang Su$^{1}$}
\email[Corresponding author. ]{Email: gsu@gucas.ac.cn}
\affiliation{$^1$College of Physical Sciences, Graduate University
of Chinese Academy of Sciences, P. O. Box 4588, Beijing 100049, China\\
$^2$College of Materials Science and Opto-Electronic Technology,
Graduate University of Chinese Academy of Sciences, P. O. Box 4588,
Beijing 100049, China}

\begin{abstract}
  A linearized tensor renormalization group (LTRG) algorithm is
  developed to calculate the thermodynamic properties of low-dimensional
  quantum lattice models. This new approach employs the
  infinite time-evolving block decimation technique, and allows for
  treating directly the transfer-matrix tensor network that makes it
  more scalable. To illustrate the performance, the thermodynamic
  quantities of the quantum XY spin chain as well as the Heisenberg antiferromagnet
  on a honeycomb lattice are calculated by the LTRG method, showing the pronounced precision
  and high efficiency.
\end{abstract}

\maketitle

Since the appearance of White's density-matrix renormalization group
(DMRG) theory \cite{White}, the numerical renormalization group (RG)
approaches have achieved great success in studying low-dimensional
strongly correlated lattice models \cite{Schollwoek}. In the past
few years, a number of RG-based methods, e.g., the coarse-graining
tensor renormalization group (TRG) \cite{Levin,Jiang,Gu}, projected
entangled pair states \cite{Cirac}, entanglement renormalization
\cite{Vidal}, the infinite time-evolving block decimation (iTEBD)
\cite{G. Vidal}, finite-temperature DMRG \cite{Verstraete,White2},
etc., \emph{have been proposed inspired by the quantum information theory}.
In spite of the great success in one- and
two-dimensional (1D and 2D) lattice models, it is still quite
necessary to develop new algorithms to improve the accuracy and
efficiency of numerical calculations for strongly correlated
systems.

In this Letter, we propose a new algorithm to simulate the
thermodynamics of low-dimensional quantum lattice models. Our
strategy is first to transform the $D$ dimensional quantum lattice
model to a $D+1$ dimensional classical tensor network by means of
the Trotter-Suzuki decomposition \cite{Trotter}, and then to
decimate linearly the tensors following the lines developed in the
iTEBD scheme to obtain the thermodynamics of the original quantum
many-body system. This algorithm is so dubbed as the linearized TRG
(LTRG). As is known, the previous real space TRG approach deals with
the 2D tensor network with exponential decimation in the
coarse-graining procedure, which was shown effective for both 2D
classical and quantum lattice models
\cite{Chang,Li1,Jiang,Gu,Chen,Li2}. For the best illustration of the
algorithm and performance of the LTRG approach, we take the exactly
solvable 1D quantum XY spin chain as a prototype. The results show
that the precision of the LTRG method is comparable with that of the
transfer-matrix renormalization group (TMRG) \cite{Xiang}, the
method that is quite powerful for simulating the 1D quantum lattice
models at finite temperatures (e.g. Refs. \cite{gucas,sirker}). To
demonstrate its scalability, a LTRG result with remarkable precision
for a 2D Heisenberg antiferromagnet on a honeycomb lattice is also
included.

\begin{figure}[tbp]
\includegraphics[angle=0,width=0.8\linewidth]{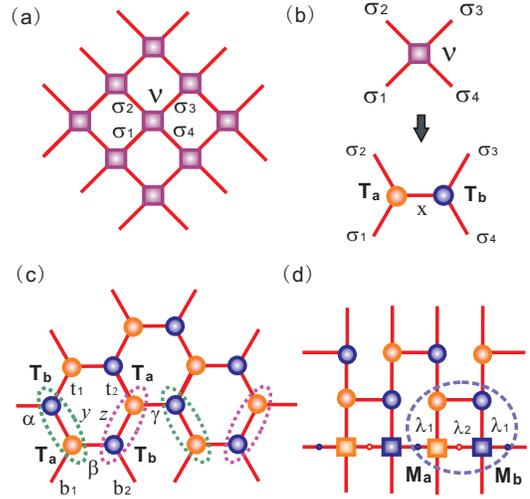}
\caption{(Color online) (a) A transfer-matrix tensor network, where
each bond denotes the $\protect\sigma $ index
in Eqs. (\protect\ref{eq-partition}) and (\protect\ref%
{eq-transfertensor}). (b) A local transformation of a fourth-order
tensor into two third-order tensors through a singular value
decomposition (SVD). (c) Transform the transfer-matrix tensor
network to a hexagonal one. (d) By contracting the intermediate
bonds marked by dashed ovals in (c), one gets a brick wall structure
with the 4th-order tensors in the bottom line.} \label{fig-LiTRG}
\end{figure}

Let us start with the Hamiltonian of a 1D quantum many-body model
given by
\begin{eqnarray} H
&=&\sum_{i=1}^{N}h_{i,i+1}=H_1+H_2,  \notag \\
H_{1}&=&\sum_{i=1}^{N/2}h_{2i-1,2i},\,\,H_{2}=\sum_{i=1}^{N/2}h_{2i,2i+1},
\label{eq-hamiltonian}
\end{eqnarray}
where $N$ (even) is the number of sites. By inserting $2K$ (large
$K$) complete sets of states $\{|\sigma_i^j\rangle\}
(\sigma_i^j=1,\cdots,D)$ with $i$ the site index and $j$ the Trotter
index, the partition function of this model can be represented as

\begin{eqnarray}
Z_{N} &\simeq & Tr[e^{-\beta H_{1}/K}e^{-\beta H_{2}/K}]^{K}  \notag \\
&=&\sum_{\{\sigma _{i}^{j}\}}\prod_{j=1}^{K}\langle \sigma
_{1}^{2j-1}...\sigma _{N}^{2j-1}|e^{-\beta H_{1}/K}|\sigma
_{1}^{2j}...\sigma _{N}^{2j}\rangle  \notag \\
&\times &\langle \sigma _{1}^{2j}...\sigma _{N}^{2j}|e^{-\beta
H_{2}/K}|\sigma _{1}^{2j+1}...\sigma _{N}^{2j+1}\rangle ,
\label{eq-partition}
\end{eqnarray}

where the periodic boundary conditions along both spatial and
temporal directions are assumed, i.e., $\sigma _{i}^{1}=\sigma
_{i}^{2K+1}$ and $\sigma _{1}^{j}=\sigma_{N+1}^{j}$. Since the terms
within $H_{1}$(and $H_2$) mutually commute, Eq. (\ref
{eq-partition}) can be further decomposed as
\begin{equation}
Z_{N}\simeq \sum_{\{\sigma
_{i}^{j}\}}\prod_{i=1}^{N/2}\prod_{j=1}^{K}v_{\sigma _{2i-1}^{2j-1}\sigma
_{2i}^{2j-1},\sigma _{2i-1}^{2j}\sigma _{2i}^{2j}}\,v_{\sigma
_{2i}^{2j}\sigma _{2i+1}^{2j},\sigma _{2i}^{2j+1}\sigma _{2i+1}^{2j+1}},
\label{eq-transfertensor}
\end{equation}
where the transfer matrix, $v_{\sigma _{1}\sigma _{4},\sigma
_{2}\sigma _{3}}\equiv\langle \sigma _{1}\sigma _{4}|\exp (-\beta
h_{i,i+1}/K)|\sigma _{2}\sigma _{3}\rangle $, is a 4th-order tensor.
Obviously, the partition function, Eq.(\ref{eq-transfertensor}), can
be viewed as a classical transfer-matrix tensor network, as
illustrated in Fig. 1(a).

\begin{figure}[tbp]
\includegraphics[angle=0,width=1.0\linewidth]{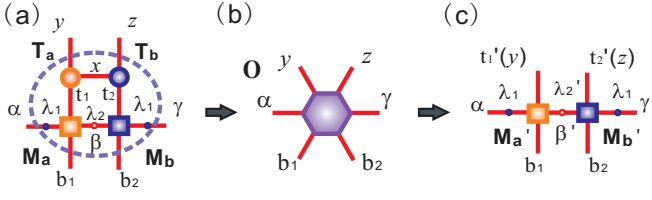}
\caption{(Color online) A local evolution of the tensors by
contraction and SVD. (a) Contract the intermediate bonds; (b) obtain
a 6th-order tensor $O$; (c) calculate the singular value
decomposition (SVD) of $O$, and update the tensors $M_{a,b}$ and
$\protect\lambda$. The above manipulation has a computational cost
that scales as $O(D^6 D_c^3)$.} \label{fig-locevol}
\end{figure}

\begin{figure}[tbp]
\includegraphics[angle=0,width=0.75\linewidth]{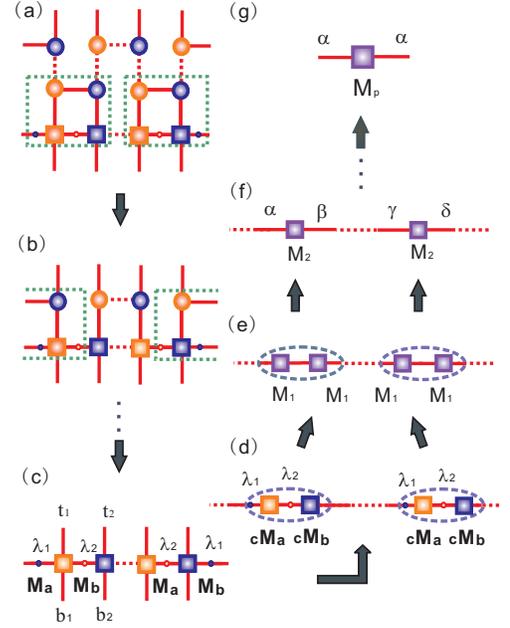}
\caption{(Color online) An successive projection of each row of
tensors onto the MPO in the bottom line [(a)-(c)]. After the
projection along the Trotter direction, by tracing out the physical
indices $t$ and $b$ of the MPO, one may get a 1D matrix product, of
which the trace can be obtained by a matrix RG procedure [(d)-(g)].}
\label{fig-matrg}
\end{figure}

The partition function can be obtained by summing over all the
intermediate states $|\sigma_i^j\rangle$, namely, contracting all the
bonds $\sigma$ in the tensor network. This procedure is accomplished
by first making a singular value decomposition (SVD) of $\nu$-tensors
in the following way
\begin{eqnarray}
\nu_{\sigma_1\sigma_2,\sigma_3\sigma_4}&=&\sum_{x=1}^{D^2}
U_{\sigma_1\sigma_2,x} \lambda_{x} V^{\top}
_{x,\sigma_3\sigma_4} \nonumber\\
&\equiv&\sum_{x=1}^{D^2}
(T_a)_{x,\sigma_1,\sigma_2} (T_b)_{x,\sigma_3,\sigma_4},
 \label{eq-TaTbsvd}
\end{eqnarray}
where the diagonal matrix $\lambda$ collects $D^2$ singular values,
and two auxiliary tensors $(T_a)_{x,\sigma_1,\sigma_2}\equiv
U_{\sigma_1\sigma_2,x}\sqrt{\lambda_x}$ and
$(T_{b})_{x,\sigma_3,\sigma_4}\equiv
V_{\sigma_3\sigma_4,x}\sqrt{\lambda_x}$ are introduced for
convenience. After this transformation, the square tensor network
becomes a hexagonal one with two 3rd-order tensors $T_{a}$ and
$T_b$, as depicted in Fig. \ref{fig-LiTRG}(b). Then, one contracts
the $\sigma$-bonds encircled by the dashed oval lines between the
last two rows in Fig. \ref{fig-LiTRG}(c), which leads to the two
4th-order tensors
\begin{eqnarray}
(M_a)_{\alpha,t_1,\beta,b_1} &=& \sum_{y=1}^{D} (T_a)_{\beta,b_1,y} (T_b)_{\alpha,t_1,y}, \notag \\
(M_b)_{\beta,t_2,\gamma,b_2} &=& \sum_{z=1}^{D} (T_a)_{\gamma,z,t_2} (T_b)_{\beta,z,b_2}, \label{eq-Ma/b}
\end{eqnarray}
which form a matrix product operator (MPO) lying in the bottom line
of the whole tensor network, that can also be viewed as a ``superket"
in the operator Hilbert space \cite{Vidal2}. Each horizontal bond
between $M_a$ and $M_b$ is assigned with a diagonal matrix
$\lambda_{1,2}$. Finally, we obtain a tensor network with brick wall
structure as shown in Fig. \ref{fig-LiTRG} (d).

Next, one can project the tensors $T_{a,b}$ onto $M_{a,b}$
successively. At each time, we project one row of tensors $T_{a}$
and $T_b$ followed by updating $M_{a,b}$ and $\lambda_{1,2}$. After
two projections, the system evolves one Trotter step forward. This
procedure is illustrated in Fig. \ref{fig-locevol}. One first
contracts the $\sigma$-bonds between $M$-tensors and $T$-tensors in
Fig.~\ref{fig-locevol} (a) to obtain a 6th-order tensor in Fig.
\ref{fig-locevol} (b)
\begin{eqnarray}
O_{y,\alpha,b_1,z,\gamma,b_2} &=& \sum_{x,t_1,t_2,\beta}
(\lambda_1)_{\alpha} \, (M_a)_{\alpha,t_1,\beta,b_1} \, (\lambda_2)_{\beta}
\, (M_b)_{\beta,t_2,\gamma,b_2} \, (\lambda_1)_{\gamma}  \notag \\
& &\,\,\,\,\,\,\,\,\,\,\,\, (T_a)_{x,t_1,y} \, (T_b)_{x,z,t_2},
\label{eq-O}
\end{eqnarray}
and then, takes a SVD of the $O$-tensors (after matricization).
$O_{y \alpha b_1,z \gamma b_2} \simeq \sum_{\beta}^{D_c} U_{y \alpha
b_1,\beta} (\lambda^{\prime}_2)_{\beta} V^{\top}_{\beta, z \gamma b_2},$
while keeps only the largest $D_c$ singular values of $\lambda'_2$.
One can define new $M$-tensors
$(M_a^{\prime})_{\alpha,y,\beta,b_1} = U_{y \alpha
b_1,\beta}/(\lambda_1)_{\alpha}$ and
$(M_b^{\prime})_{\beta,z,\gamma,b_2} = V_{z \gamma
 b_2,\beta}/(\lambda_1)_{\gamma}$, and update the horizontal
bonds with $\lambda^{\prime}_2$.  After these operations, the last
row of the tensor network is half updated as shown in Fig.
\ref{fig-locevol} (c).  To project the next row of tensors, one can
simply exchange $M_{a}$ and $M_{b}$ as well as $\lambda_{1}$ and
$\lambda_2$ in Eq. (\ref{eq-O}). These two successive projections
make up of a full Trotter step $\tau$, as illustrated from Fig.
\ref{fig-matrg}(a) to \ref{fig-matrg}(c). In each Trotter step, the
transfer-matrix tensor network is decimated linearly with  only
$O(D_c)$ singular values discarded, which improves greatly the
efficiency compared with the original TRG approach where $O(D_c^n)$
($n=2$ for honeycomb network) ones are discarded in the
coarse-graining procedure \cite{expla}.

In order to avoid the divergence in the imaginary time evolution,
one has to normalize all the singular values in $\lambda$ with its
largest one $n_i$ in $i$-th step. After projecting all the
$T$-tensors at inverse temperature $\beta$, one is left with the
matrix product density operator of the present system. It consists
of 4th-order $M$-tensors [see Fig. \ref{fig-matrg} (c)], each of
which has two legs with physical indices $t$ and $b$ in the Trotter
direction, that can be further traced out due to the periodic
boundary condition. Thus, we obtain a 1D matrix product (MP)
extended in the spatial direction, where the matrices are labeled as
$cM_{a,b}$ as shown in Fig. \ref{fig-matrg} (d). It is convenient to
assume the number of matrices is $2^p$. To get the trace of the
product of these $2^p$ matrices, one can contract the neighboring
matrices pairwise to obtain a new product of $2^{p-1}$ matrices,
each of which should be normalized by the absolute value of its
largest elements to avoid divergence. This contraction procedure is
represented in Figs. \ref{fig-matrg}(d)-\ref{fig-matrg}(g). After
$p$ steps, the $2^p$ matrices shrink to a single one, of which the
trace can be easily calculated. In each coarse graining step, all
the normalization factors denoted by $m_j$ with $j=1, \cdots,p$ need
to be collected for the calculation of physical quantities, e.g.,
the free energy per site $f$ at inverse temperature $\beta=K\tau$
can then be determined by the normalization factors $n_j$'s and
$m_j$'s
\begin{eqnarray}
f & = & -\frac{1}{\beta L} \ln[ \prod_{i=1}^{2K-2}(n_i)^{\frac{L}{2}}
\prod_{j=1}^{p}(m_j)^{\frac{L}{2^{j}}} ]  \notag \\
& = & -\frac{1}{K\tau}(\sum_{i=1}^{2K-2}\frac{\ln{n_i}}{2}+\sum_{j=1}^{p}%
\frac{\ln{m_j}}{2^{j}}).  \label{eq-feng}
\end{eqnarray}

\begin{figure}[tbp]
\includegraphics[angle=0,width=0.85\linewidth]{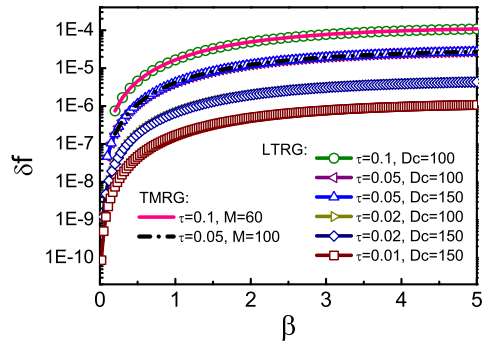}
\caption{(Color online) The relative error of the free energy per
site, $\protect\delta f$, of the quantum XY spin chain at high
temperatures. $\delta f$ converges rapidly with $D_c$, and the lines
with $D_c=100$ and $150$ coincide with each other
($\tau=0.05,0.02$). In addition, the TMRG results ($\tau=0.1,0.05$)
are also presented for a comparison.} \label{fig-relaerror}
\end{figure}

In the above descriptions, we illustrate the LTRG algorithm by first
decimating the tensors along the Trotter direction, and then
contracting the matrices in the spatial direction. Alternatively,
one can also perform the decimation first in the spatial direction,
and then do the matrix contraction in the Trotter direction.

As an example, we are going to demonstrate the efficiency of the
LTRG algorithm by computing the free energy and other thermodynamic
quantities of the quantum XY spin-1/2 chain with a local Hamiltonian
$h_{i,i+1}= -J(S_{i}^{x}S_{i+1}^{x}+S_{i}^{y}S_{i+1}^{y})$ in Eq.
(\ref{eq-hamiltonian}) with $J=1$. We take the chain length to be
$2^{100}$, which definitely reaches the thermodynamic limit.

\begin{figure}[tbp]
\includegraphics[angle=0,width=1.05\linewidth]{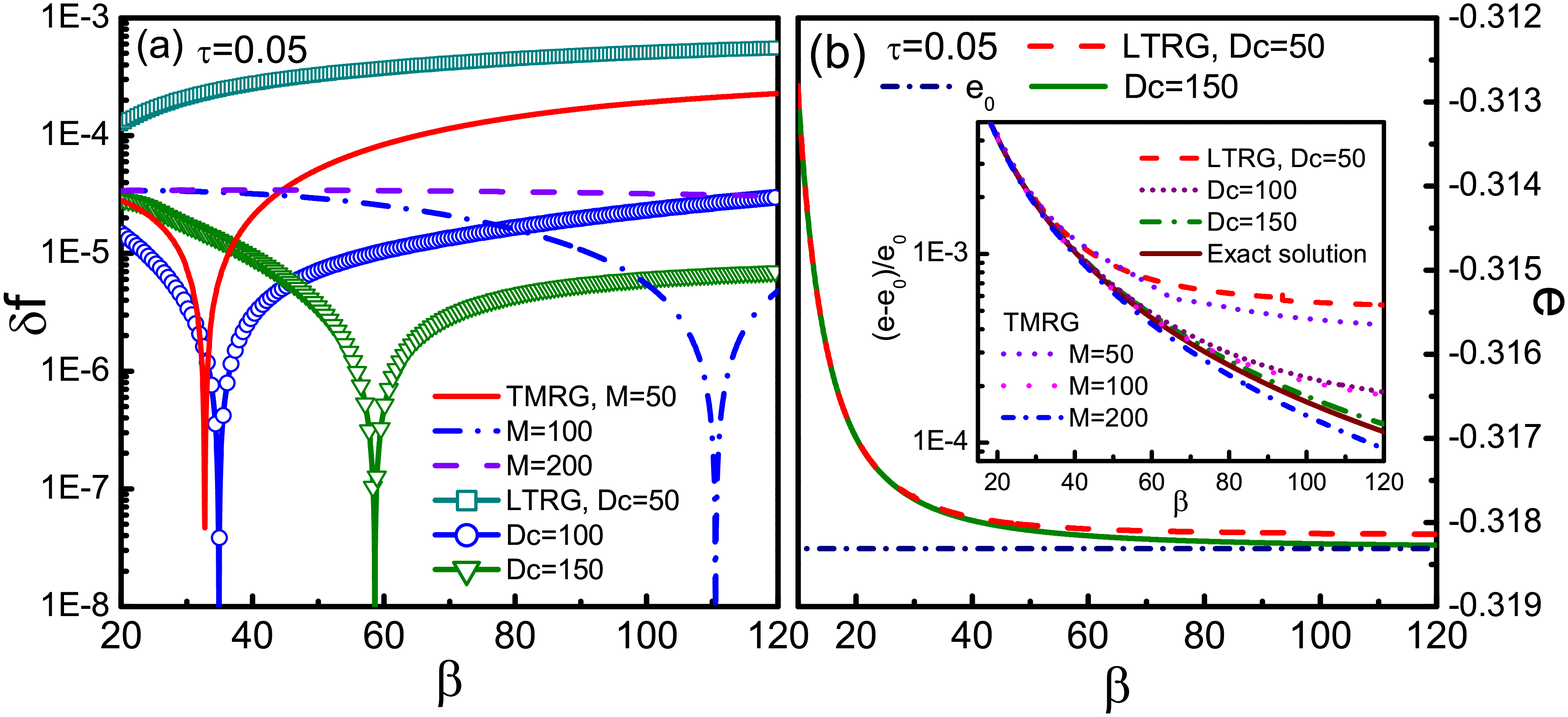}
\caption{(Color online) LTRG and TMRG results of the quantum XY spin
chain. (a) Relative error of the free energy per site
$\protect\delta f$. (b) The energy per site $e$. The inset shows the variation of $%
(e-e_0)/e_0$ with inverse temperature $\beta$ for various $D_{c}$.}
\label{fig-lowT}
\end{figure}

In Fig. \ref{fig-relaerror}, we show the relative error of the free
energy $f$ with respect to the exact solution, i.e., $\delta f=
|(f-f_{exact})/f_{exact}|$, for different Trotter steps $\tau=0.1,
0.05, 0.02, 0.01$. We observe that the accuracy is enhanced with
decreasing $\tau$, as well as increasing $D_c$. Owing to the close
relation between iTEBD and DMRG, the truncation parameter $D_c$
plays a role similar to the number of states kept $M$ in the TMRG
method. As shown in Fig. \ref{fig-relaerror}, we compare the LTRG
results to those of TMRG, both of which show the same accuracy for
$\tau=0.1$ and $0.05$. It is also noticed that the relative errors
saturate rapidly with increasing $D_c$, implying that the errors at
high temperatures (\emph{e.g.} $T > 0.2J$) mainly originate from the
Trotter-Suzuki decomposition. In order to check the truncation
error, the LTRG algorithm is also tested at very low temperatures.
In Fig. \ref{fig-lowT}, the temperature is down to $T=J/120$ with a
Trotter step $\tau=0.05$. As shown in Fig. \ref{fig-lowT} (a), the
accuracy of low $T$ results is remarkably improved by increasing
$D_c$, and the relative error $\delta f\simeq7\times 10^{-6}$ at
$\beta=120$ for $D_c=150$.

Besides the free energy, other thermodynamic quantities, such as the
internal energy, can also be obtained. There are at least two ways
to get them, one can either introduce some impurity tensors in the
tensor network (see, for instance, Ref. \onlinecite{G. Vidal}), or
do a numerical differentiation of free energy with respect to
temperature. Both ways are found to have a similar accuracy. In Fig.
\ref {fig-lowT} (b), the energy per site, $e$, is presented. We
apply the LTRG algorithm to approach the ground state energy $e_0$,
and find the difference $(e-e_0)/e_0$ is about $10^{-4}$ at $
\beta=120$ for $D_c$=150, suggesting that the LTRG result is very
close to the exact solution. The TMRG results with various $M$ (up
to $M=200$) are also included in Fig. \ref{fig-lowT} for a
comparison. The relative errors for the free energy and internal
energy are found to be of the same order down to $\beta=120$ for
both approaches.

\begin{figure}[tbp]
\includegraphics[angle=0,width=1.05\linewidth]{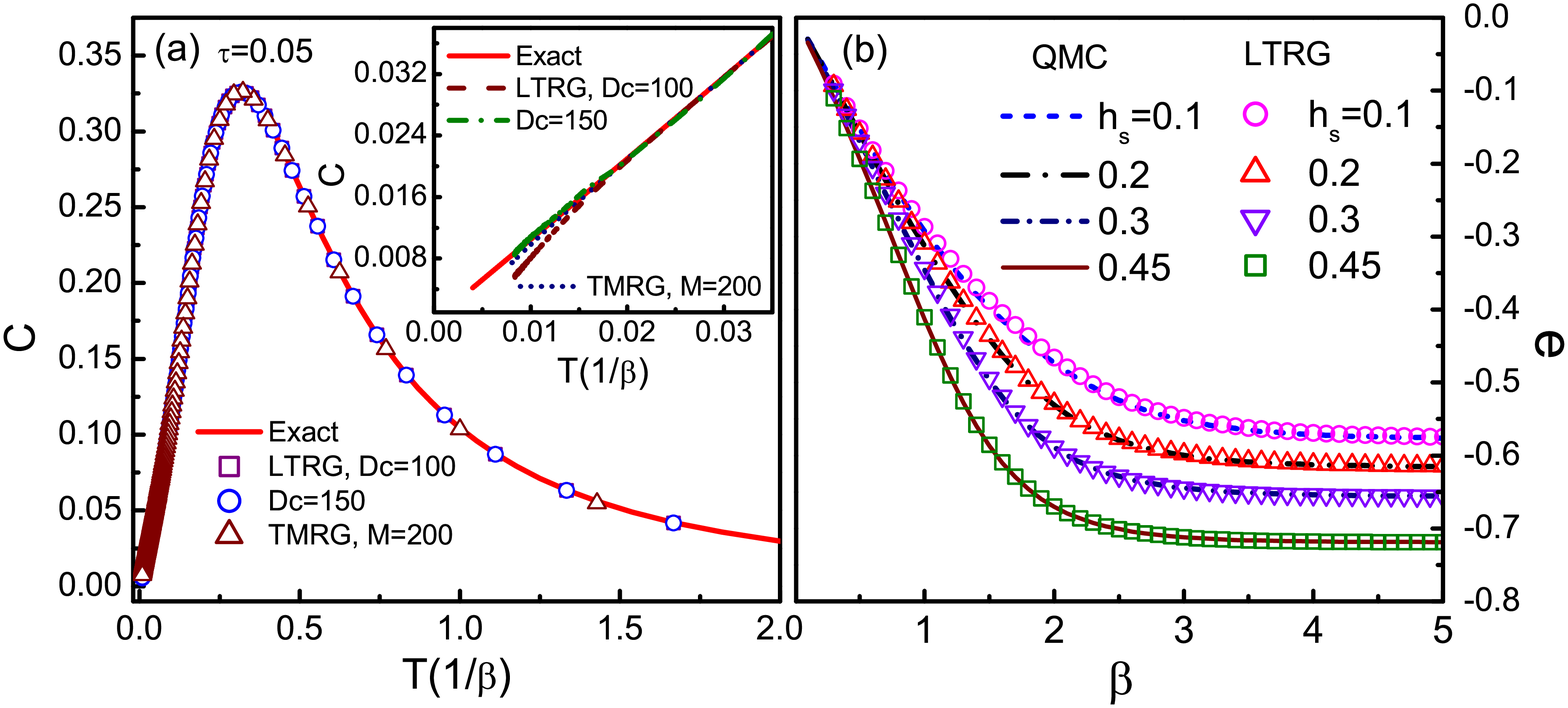}
\caption{(Color online) (a) Specific heat as a function of
temperature ($T=1/\beta$) of the quantum XY spin chain. The inset
shows the low temperature results for $D_{c}=100$ and $200$, along
with the TMRG data ($M=200$) for a comparison. (b) Energy per site
of the 2D spin-1/2 Heisenberg antiferromagnet on a honeycomb lattice
for different staggered magnetic fields. The QMC results are
obtained by using the ALPS library \cite{AF}.} \label{fig-speC}
\end{figure}

The specific heat of the quantum XY spin chain is also calculated,
as shown in Fig. \ref{fig-speC}(a). The LTRG results agree quite
well with the exact solution both at high and low temperatures. As
indicated in the inset, the accuracy will be enhanced by increasing
$D_c$. For $D_c=150$, the LTRG results coincide with the exact
solution down to very low temperature ($T/J\simeq0.008$). The TMRG
results with states $M=200$ are also included, showing that both
numerical methods have the comparable accuracy.

To examine the scalability of the LTRG algorithm, we also calculate
the energy per site of a 2D spin-1/2 Heisenberg antiferromagnetic
model on a honeycomb lattice, whose Hamiltonian is $H=J
\sum_{<i,j>}\vec{S}_i \cdot \vec{S}_j + h_s
\sum_{i}(-1)^{|i|}S_i^z$, where $(-1)^{|i|}$ denotes the parity of
the lattice and $h_s$ is a staggered magnetic field, as shown in
Fig. \ref{fig-speC} (b). A pronounced agreement between LTRG and
quantum Monte Carlo (QMC) results is clearly seen.

In summary, we have proposed a linearized TRG algorithm to calculate
the thermodynamic properties of low dimensional quantum lattice
models, and obtained very accurate results. The LTRG algorithm can
be readily generalized to fermion and boson models, and also
provides a quite promising way to simulate the 2D quantum lattice
models without involving the negative sign problem.

We are indebted to Q. N. Chen, J. W. Cai, J. Sirker, T. Xiang, Z. Y. Xie, and H. H. Zhao
for stimulating discussions, and Z. Y. Chen, S. J. Hu, Y. T. Hu, G. H. Liu, X. L. Sheng, Y. H. Su,
Q. B. Yan, and Q. R. Zheng for helpful assistance. This work is supported in part by the NSFC
(Grants No. 10625419, No. 10934008, No. 10904081, No. 90922033) and the Chinese Academy of Sciences.


\begin{thebibliography}{99}
\bibitem{expla} We note that the present LTRG algorithm can also be applied to
evaluate the thermodynamics of 2D classical models, achieving more
accurate results than the coarse-graining TRG algorithm.
\end{thebibliography}
\end{document}